# Compact Model of a Topological Transistor

Md Mazharul Islam[1], Shamiul Alam[1], Md Shafayat Hossain[2], and Ahmedullah Aziz[1*]
[1]Dept. of Electrical Eng. & Computer Sci., University of Tennessee, Knoxville, TN, 37996, USA
[2]Dept. of Physics, Princeton University, Princeton, NJ, 08544, USA

**Abstract**

**The precession of a ferromagnet leads to the injection of spin current and heat into an adjacent non-magnetic material. Besides, spin-orbit entanglement causes an additional charge current injection. Such a device has been recently proposed where a quantum-spin hall insulator (QSHI) in proximity to a ferromagnetic insulator (FI) and superconductor (SC) leads to the pumping of charge, spin, and heat. Here we build a circuit-compatible Verilog-A-based compact model for the QSHI-FI-SC device capable of generating two topologically robust modes enabling the device operation. Our model also captures the dependence on the ferromagnetic precision, drain voltage, and temperature with an excellent (> 99%) accuracy.**

The advent of transistors has revolutionized the progression of human civilization in an unimaginable way. The consistent miniaturization of the transistor in the last few decades has made us capable of storing and processing vast amounts of data [1]. However, the quantity of data to be processed has also been increasing exponentially side by side[2]. While this dimensional reduction has taken place, researchers have already arrived at the theoretically predicted physical bottleneck[3–5]. In this context, topological materials stand in the spotlight for the exploration of future low-power and robust computational devices[6,7]. The gapless surface state in topological insulators has made it a focal point of research[6,8]. Although the technology is still in a nascent stage, there have been numerous remarkable efforts with significant research findings[9,10]. Among them, there have been several propositions of the topological phenomenon-driven device structures[10–16]. Becerra *et. al.* has recently proposed such a device where a quantum spin Hall insulator (QSHI) adjacent to a ferromagnetic insulator (FI) and a superconductor (SC) can harbor Majorana zero mode in the FI-SC junction[17]. This Majorana fermion (MF) enables topologically protected perfect Andreev reflection (AR)[18]. The precessing magnetization of the FI region enables interrelated and quantized spin, heat, and charge pumping. Here, the MF enables two unique topological operation regimes where the pumping of electrons can be turned on and tuned by external control parameters. Also, the injection of heat, spin, and charge are exponentially sensitive to the external control parameters (gate voltage, precession angle) similar to the conventional transistor behavior. The device has two operation regimes that are topologically protected offering immunity to disorder and other imperfections. Moreover, the device offers sufficient scalability as QSHI can be patterned with FI and SC by various deposition methods [19–21]. So, a physics-informed compact model can serve as a handy tool to explore and generate valuable insights about the device that will leverage future research endeavors.

In this work, we develop a physics-informed compact model for the QSHI-FI-SC device. We simplify the rigorous mathematical expression of the injected spin, heat, and charge to a closed-form equation suitable for compact modeling. The piecewise linear approximation method is used to reduce the mathematical complexity. The deduced expression captures the dependency of the heat, spin, and charge injection on the external control parameters with great precision. Using the deduced closed form equation, we build a Verilog-A-based compact model. Our model can perfectly capture the proposed transistor behavior of the two topological regimes reported in [17]. Our model will enable the device and circuit-level exploration of the device. Besides, our modeling approach provides a pathway for future device modeling with a high level of mathematical complexity.

To describe the system, the Bogoliubov-de Gennes Hamiltonian is considered as,

$$H_{BdG}(t) = [v_F p\sigma_z - \mu(x)]\tau_z + m(x,t).\sigma + \Delta(x)\,\tau_z\,, \tag{1}$$

where $\sigma = (\sigma_x, \sigma_y, \sigma_z)$ and $\tau = (\tau_x, \tau_y, \tau_z)$ are the Pauli matrices acting on the spin and Nambu space. $v_F$ is the Fermi velocity and $m(x,t)$ is the time-dependent magnetization of the FI region. $\Delta(x)$ and $\mu(x)$ represent the superconducting order parameter and chemical potential, respectively. Throughout the whole SC region, $\Delta(x)$ is assumed to be constant as $\Delta_0$. The $m(x,t)$ is periodically driven in the FI region. This results in the pumping of the charge, spin, and heat in the left lead of the device (Fig. 1). The value of magnetization is parameterized as $m(x,t) = m_0(x)\,[sin\,\theta(t).cos\,\phi(t), sin\,\theta(t).sin\,\phi(t), cos\,\theta(t)]$ where $m_0(x)=m_0$ is the magnetization in the FI region . In the scattering matrix formalism, the only nonzero reflection co-efficient are corresponding to the normal and Andreev reflections. Here, $r^{\uparrow\downarrow}_{he(eh)}$ $(E, \theta, \phi)$ represent the reflection amplitude for the electron (hole) with spin ↓ injected from the QSHI to be reflected as a hole (electron). Here, $r^{\uparrow\downarrow}_{ee(hh)}$ $(E, \theta, \phi)$ represent the reflection amplitude of the electron (hole) that has a spin ↓ and injected from the QSHI to be reflected as an electron (hole). These coefficients are related as $|r^{\uparrow\downarrow}_{ee(hh)}(E, \theta, \phi)|^2 + |r^{\uparrow\downarrow}_{he(eh)}(E, \theta, \phi)|^2 = 1$. The reflection coefficients can be expressed as, $r^{\uparrow\downarrow}_{ee(hh)}(E, \theta, \phi)$ $= r_0(E, \theta)\, e^{i\phi}$ and $r^{\uparrow\downarrow}_{hh(ee)}(E, \theta, \phi) = -\,[r^{\uparrow\downarrow}_{ee(hh)}(-E,\theta,\phi)]^*$. At a sufficiently low energy, the magnitude of the coefficient representing the normal reflection is suppressed due to the perfect AR. In other words, at low energy, $|r^{\uparrow\downarrow}_{he(eh)}(E{=}0, \theta, \phi)| = 1$, at low energy, the $\phi$ independent part can be approximated as

$$|r_0(E,\theta)|^2 \approx \frac{E^2/\Gamma^2}{1+E^2/\Gamma^2} \tag{2}$$

Here, Γ is the Majorana linewidth for which the normal and Andreev coefficients are equal. Γ is defined as-

$$\Gamma = 2\Delta_0 (\frac{\xi_F\,(0,\theta)}{\xi_F\,(V_g,\theta\,)})^2 \frac{\xi_S}{\xi_F\,(V_g,\theta) + \xi_S}\, e^{-\frac{2L}{\xi_F(V_g,\theta)}} \tag{3}$$

$\xi_F$ and $\xi_S$ are the coherence length of FI and SC region respectively. These can be expressed as, $\xi_F\,(V_g,\theta) = \frac{\hbar v_F}{\sqrt{(m_0^2 \sin^2\theta - (eV_g)^2}}$ and $\xi_s\,(V_g,\theta) = \frac{\hbar v_F}{\Delta_0}$. The chemical potential of the FI region is controllable by the gate voltage ($V_g$) and is denoted as $\mu_{FI} = eV_g$. The length of the FI region is represented by $L$. The injection of charge, spin, and heat pumped in a single cycle can be described by a single parameter. This is referred as the dimensionless charge ($\mathcal{Q}$) and is related to the charge, spin and heat as $Q_e = e\mathcal{Q}$, $S_Z = -\frac{\hbar}{2}\mathcal{Q}$, and $Q_E = -\frac{\hbar\omega}{2}\mathcal{Q}$, respectively. The dimensionless charge $\mathcal{Q}$ can be expressed as:

$$\mathcal{Q} = -\frac{1}{2\pi}\int dE\,(\frac{\partial f(E)}{\partial E})\int_0^{2\pi} d\phi\,|r_{ee}^{\downarrow\uparrow}\,(E,\theta,\phi)|^2 \tag{4}$$

Thus, it depends on the Fermi energy ($E_F$) via the fermi function [$f(E)$] and the reflection coefficient ($r_{ee}^{\downarrow\uparrow}$). $\mathcal{Q}$ is sensitive to the temperature ($T$) and the applied drain voltage ($V_d$) via the Fermi function $f(E)$. It is also dependent on the rotation angle ($\theta$) and the gate voltage ($V_g$) via $|r_{ee}^{\downarrow\uparrow}|^2$.

Now, taking the energy derivative of the Fermi energy and integrating $|r_{ee}^{\downarrow\uparrow}\,(E,\theta,\phi)|^2$, in the adiabatic limit, the dimensionless charge can be expressed as

$$\mathcal{Q} = \int \frac{E^2 e^{\frac{E-(E_F+eV_d)}{kT}}}{kT\,(E^2+\Gamma^2)\left(e^{\frac{E-(E_F+eV_d)}{kT}}+1\right)^2} dE \tag{5}$$

The internal dynamics between the QSHI, FI, and SC are merged in equation (5) in an integral format. The integral expression is indefinite and is required to be expressed in a compact format in order to develop a circuit compatible Verilog-A model. To do so, we approximate the exponential component of the integrad. For simplicity, we are defining three different parameters as $\alpha = eV_d + E_F$, $\beta = kT$, $\gamma = \Gamma$. Now, the bell-shaped exponential component of equation (5) can be approximated as a piecewise linear function as below:

$$\frac{e^{\frac{E-\alpha}{\beta}}}{\left(e^{\frac{E-\alpha}{\beta}}+1\right)^2} = \begin{cases} 0 & ;\; E < -3.8\beta + \alpha \\ 0.07\,\frac{E-\alpha}{\beta} + 0.2660\,; & -3.8\beta + \alpha < E < 0 \\ -\,0.07\frac{E-\alpha}{\beta} + 0.2660; & 0 < E < 3.8\beta + \alpha \\ 0 & ;\; E > 3.8\beta + \alpha \end{cases} \tag{6}$$

Now, $\mathcal{Q}$ over all possible energy range can be approximated as the following,

$$\mathcal{Q} = \int_{-3.8\beta+\alpha}^{0} \frac{E^2}{\beta(E^2+\gamma^2)}\left(0.07\frac{E-\alpha}{\beta} + 0.2660\right) dE + \int_{0}^{3.8\beta+\alpha} \frac{E^2}{\beta(E^2+\gamma^2)}\left(-0.07\frac{E-\alpha}{\beta} + 0.2660\right) dE \tag{7}$$

Now, the result of the definite integral describes the closed form expression of the dimensionless charge as below,

$$\mathcal{Q} = \frac{1}{5000\beta^2}[175\gamma^2 . ln\left(\frac{25\gamma^2+361\beta^2-190\alpha\beta+25\alpha^2}{25(\gamma^2+\alpha^2)}\right) + (350\alpha\text{-}1330\beta). \gamma.[tan^{-1}(\frac{\alpha}{\gamma})+tan^{-1}(\frac{19\beta-5\alpha}{5\gamma})]$$

$$+175\gamma^2 . ln\left(\frac{25\gamma^2+361\beta^2+190\alpha\beta+25\alpha^2}{25(\gamma^2+\alpha^2)}\right) + (350\alpha+1330\beta).\gamma.[tan^{-1}(\frac{\alpha}{\gamma})\text{-}tan^{-1}(\frac{19\beta+5\alpha}{5\gamma})] + 2\times2527\beta^2] \tag{8}$$

We calibrate our model with the geometric and material parameters corresponding to the QSHI-FI-SC structure reported in [17] (Table I). Using equation (8), we build a circuit-compatible Verilog-A-based compact model. We validate our model with the device simulation result from [17]. Figure 2 shows the dependence of $\mathcal{Q}$ on $V_d$ keeping all the other parameters at their nominal values. The device transforms from the topological suppression regime to the quantization regime as $V_d$ goes from low to high voltage. It is evident that, the model can capture the basic transistor behavior and matches with the device simulation results. Furthermore, the impact of $V_g$ and $\theta$ on $\mathcal{Q}$-$V_d$ characteristics is also portrayed perfectly by our model (Figs. 2 (a) and (b), respectively). For a certain $V_d$, $\mathcal{Q}$ decreases for higher $V_g$ and increases with $\theta$. Figure 3 shows the temperature dependence of $\mathcal{Q}$. Here, $\mathcal{Q}$ increases with $T$ and transfers from one (suppression) to another (quantization) topological regime as $T$ increases. For a certain $T$, $\mathcal{Q}$ decreases with $V_g$ and $\theta$, and our model perfectly matches with the reported device simulation result (Figs. 3 (a) and (b), respectively). We also examine the variation of $\mathcal{Q}$ with respect to the ferromagnetic precision frequency ($\omega$) as shown in Fig. 4. At lower $\omega$, the device operates in the topological suppression regime and $\mathcal{Q}$ has a negligible value. On the contrary, at higher $\omega$, $\mathcal{Q}$ has a high value owing to the high injection of spin current at the left lead. The excellent agreement between the theory and simulation results attests that our model can capture the $\omega$ variation as well. However, our model is limited in scope as it does not converge at 0 K, contrary to the theoretically demonstrated device simulation. However, our model can describe the characteristics of $\mathcal{Q}$ for any other value of $T$ and all possible values of $\omega$ and $V_d$. We set a miniscule nominal value of $T$ (0.1 mK) for our simulation. From Figs. 2-4, it can be said that our model matches with the theoretically demonstrated device simulation behaviors. The worst-case mismatch between the reported device simulation data and our model is <0.5%.

In summary, we developed a compact model in Verilog-A for a QSHI-FI-SC structure via deriving a closed-form expression of the injected spin, heat, and charge by a piecewise linear approximation method. The model behavior is benchmarked with the device simulation to verify its functional behavior. Our Verilog-A model enables the future exploration of the circuit and system-level applications of the device.

**Table I: Device parameters for the QSHI-FI-SC structure**

| Parameters | Values |
|---|---|
| Superconducting order parameters ($\Delta_0$) | 1 meV |
| Magnetization ($m_0$) | 2 meV |
| Chemical potential ($\mu_0$) | $eV_g$ |
| Fermi Velocity ($\nu_F$) | $2.7 \times 10^7$ ms$^{-1}$ |
| Length of the Ferromagnetic insulator (FI) region ($L$) | 400 nm |

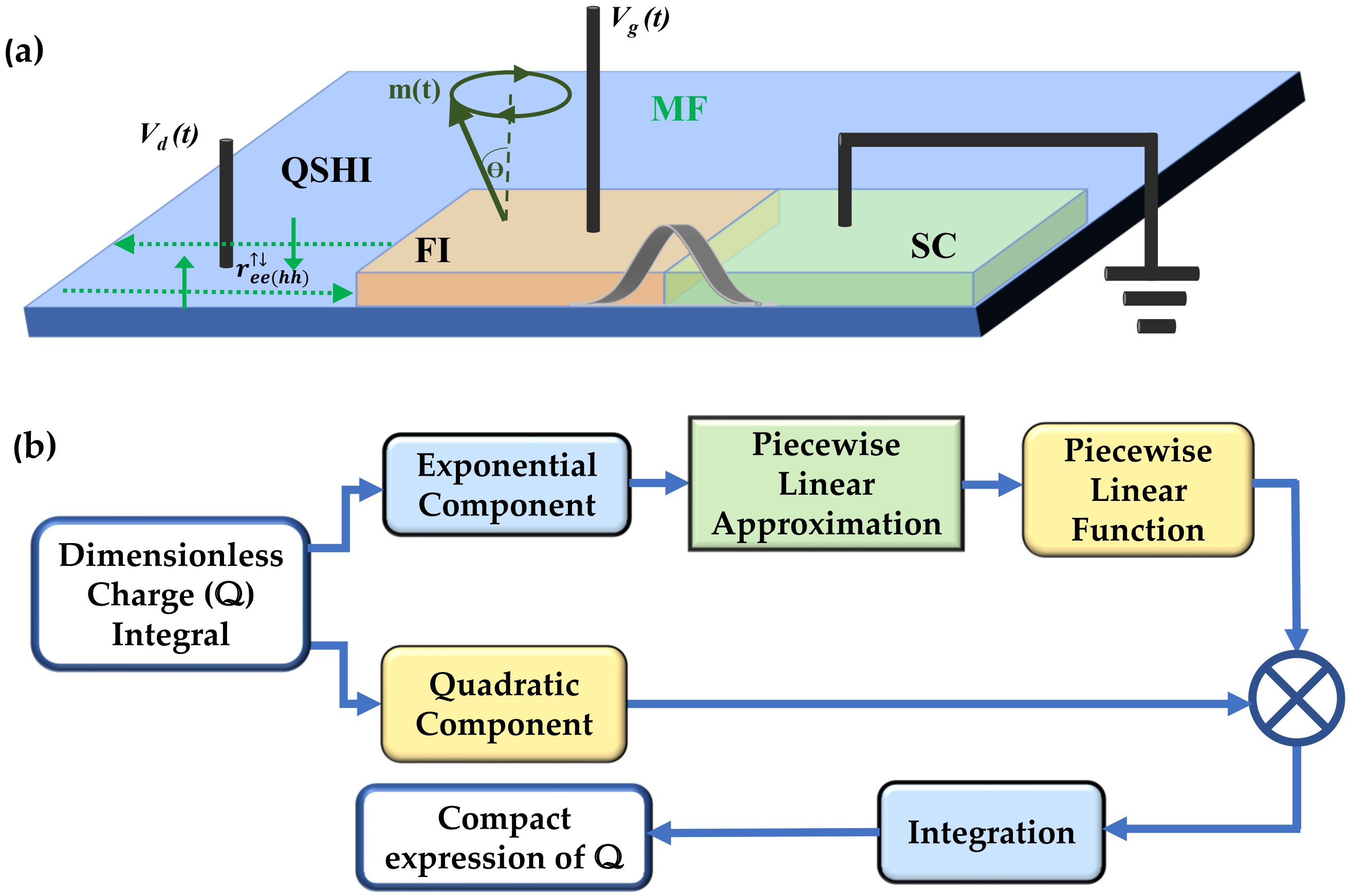

**Fig. 1: (a)** Schematic structure of a topological transistor. A QSHI region in proximity with the FI with a monodomain magnetization m(t) that precesses at an angle $\theta$. In proximity to the FI region there is a SC region The monodomain magnetization *m(t)* precesses at an angle $\theta$ around the axis perpendicular to the QSHI. The QSHI region injects charge, spin, and heat currents to the drain. The injection can be tuned by the applied potential at the FI region (*Vg*), the precession angle (θ), precession frequency ($\omega$) temperature (T) and drain voltage ($V_d$). Zero energy Majorana Fermion (MF) is harbored in the FI-SC interface that controls the pumped currents. **(b)** Methodology flow for compact modeling.

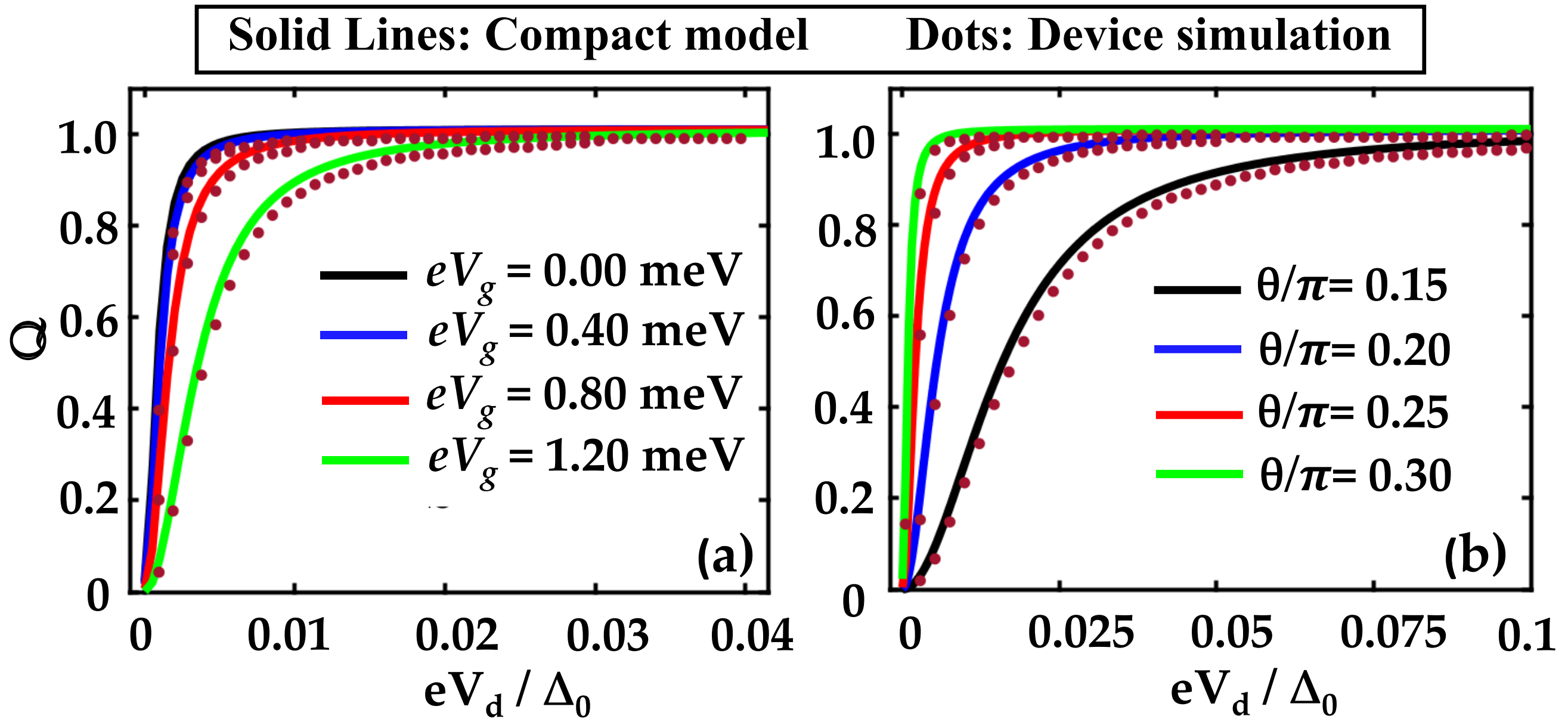

**Fig. 2:** Drain voltage dependence of the dimensionless charge ($\mathbb{Q}$). The compact model characteristics are plotted alongside with the device simulation data. Compact model (solid lines) accurately captures the datapoints from the device simulation (dotted brown). The dimensionless charge $\mathbb{Q}$ in the adiabatic limit as a function of $V_d$ for various values of **(a)** $V_g$ and **(b)** $\theta$.

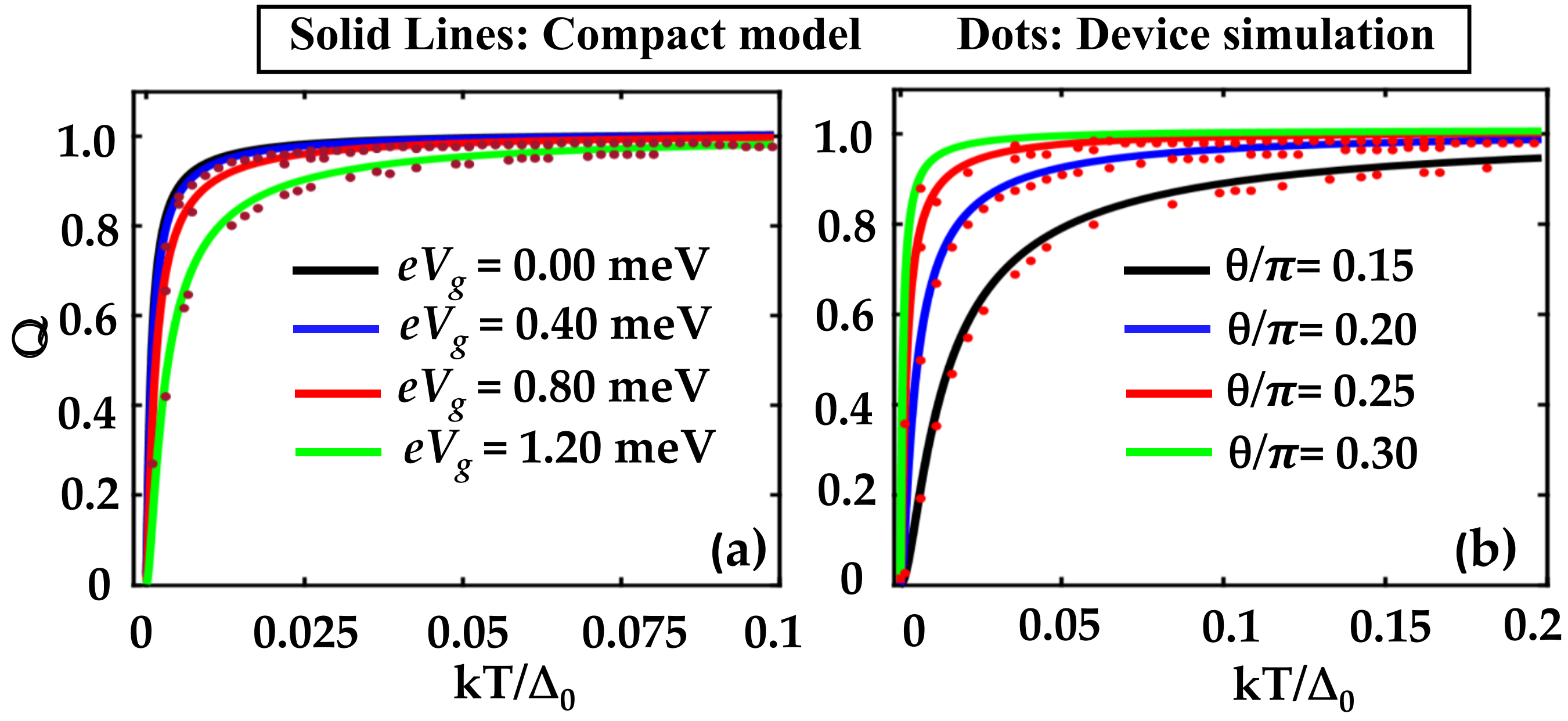

**Fig. 3:** Temperature dependence of the dimensionless charge ($\mathbb{Q}$). The compact model characteristics are plotted alongside with the device simulation data. Compact model (solid lines) accurately captures the datapoints from the device simulation (dotted brown). The dimensionless charge $\mathbb{Q}$ in the adiabatic limit as a function of $T$ for various values of **(a)** $V_g$ and **(b)** $\theta$.

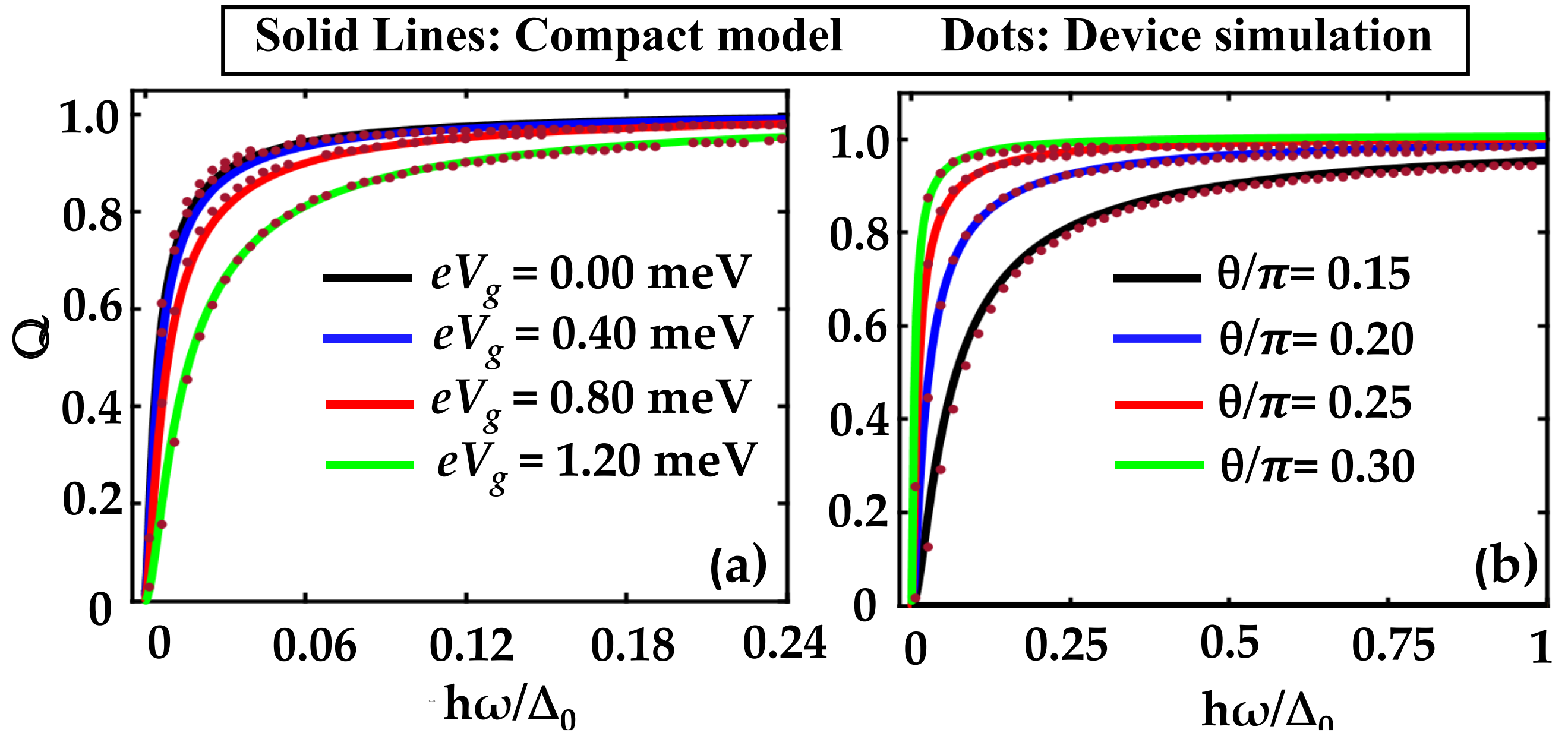

**Fig. 4:** Precession frequency dependence of the dimensionless charge ($\mathbb{Q}$). Compact model (solid lines) accurately captures the datapoints from the device simulation (dotted brown). The dimensionless charge $\mathbb{Q}$ in the adiabatic limit as a function of $\omega$ for various values of **(a)** $V_g$ and **(b)** $\theta$.

**Data Availability**

The data that support the plots within this paper and other finding of this study are available from the corresponding author upon reasonable request.

**Author Contribution**

M.M.I. conceived the idea and developed the compact model. S.A. performed the approximation method. M.S.H. and A.A. analyzed the results. All authors commented on the results and wrote the manuscript. A.A. supervised the project.

**Competing Interest**

The authors declare no competing interests.